\documentclass[doublecol]{epl2}

\usepackage{bbold,graphicx,bm,amsmath,amssymb}

\newcommand{\w}{\textrm{w}}
\newcommand{\s}{\textrm{s}}
\newcommand{\W}{\textrm{W}}
\renewcommand{\S}{\textrm{S}}
\renewcommand{\u}{\textrm{u}}
\renewcommand{\v}{\textrm{v}}

\renewcommand{\d}{\textrm{d}}

\title{Noise-driven bias in the non-local voter model}

\author{Kevin Minors\inst{1}, Tim Rogers\inst{1}, Christian A Yates\inst{1,2}}
\institute{                    
  \inst{1} Centre for Networks and Collective Behaviour\\
  \inst{2} Centre for Mathematical Biology\\
  Department of Mathematical Sciences, University of Bath, Bath, BA2 7AY, UK
}


\abstract{Is it more effective to have a strong influence over a small domain, or a weaker influence over a larger one? Here, we introduce and analyse an off-lattice generalisation of the voter model, in which the range and strength of agents' influence are control parameters. We consider both low and high density regimes and, using distinct mathematical approaches, derive analytical predictions for the evolution of agent densities. We find that, even when the agents are equally persuasive on average, those whose influence is wider but weaker have an overall noise-driven advantage allowing them to reliably dominate the entire population. We discuss the implications of our results and the potential of our model (or adaptations thereof) to improve the understanding of political campaign strategies and the evolution of disease.}

\begin{document}

\maketitle

\section{Introduction}

The voter model \cite{Holley1975,Liggett2013} and its many variants provide a highly simplified framework through which to explore the possible behaviours of a wide range of emergent natural phenomena. The myriad of applications of models of this class include: competing populations in biology \cite{kimura1964ssm,crow1970ipg,Clifford1973,hallatschek2007gde,pigolotti2014sad}; the decomposition of alloys and the kinetics of heterogeneous catalysts in physics \cite{bray2002tpo,frachebourg1996erk,krapivsky1992kmm}; and, of course, opinion dynamics in social and economic contexts \cite{fernandez2014vmm,fortunato2007sup,castellano2009sps}. The voter model was originally posed as a lattice-based interacting particle system in which each of $N$ individuals periodically re-evaluates their opinion (represented as an element of $\{0,1\}$), selecting their new opinion at random from those of their neighbours \cite{Holley1975}. This simple dynamical rule leads to a striking \emph{coarsening} phenomenon in which domains of homogeneous opinion grow over time until eventually (after a simulation time proportional to $N^2\log N$ \cite{Cox1989}) a large random fluctuation breaks the symmetry between opinions, leading to one or other managing to take over the whole domain --- a state referred to as \emph{consensus}. 

Over the decades since its conception, a large body of theory has developed exploring how variations to the voter model specification can affect the process of consensus formation. Notable examples include the addition of memory\cite{Stark2008}, conservation laws \cite{Caccioli2013}, dynamic network structure\cite{Holme2006,Ehrhardt2006,Rogers2013}, population growth \cite{Morris2014}, off-lattice agent dynamics \cite{pigolotti2014sad,pabjan2008mof}, opinion dynamics in Brownian particles \cite{schweitzer2007brownian}, and many more. However, amongst these many variants an important avenue has remained unexplored. In voter-type models it is common to assume that agents interact only with their immediate neighbours, and that each agent type is equally likely to convert the other. Thinking about the origins of the model as an analogy of political dynamics, however, it is clear that in the real world activists face a choice in how to target their efforts: should they spend all their energies convincing their immediate neighbours, or is it better to spread their message more widely, but with reduced intensity?

To address this question, we propose a simple, novel model of spatial consensus-forming dynamics, that is a non-local and off-lattice generalisation of the voter model. We consider a population composed of agents of two types, labelled $\W$ and $\S$, holding differing opinions and with different strategies for influencing others. The agents move according to Brownian motions, and attempt to convert other agents of the opposite type that are within their conversion radius. Crucially, we consider the case that the conversion radii and probability of successful conversion are unequal, with the type $\W$ agents having a wide conversion radius but converting weakly, and the type $\S$ agents having a small radius but strong power of conversion. Fixing the product of conversion strength and the size of the region of conversion ensures that the two types of agent are equally persuasive on average. Simulation outcomes are not equally balanced, however: in large populations we find that the wide and weak ($\W$-type) agents are almost certain to achieve consensus. 

In this letter we present a theoretical analysis of our non-local voter model to elucidate the origins of the bias in favour of the wide and weak persuasion strategy. We find that, in fact, different mechanisms are at work depending on the density of agents. In the low-density regime (i.e. when the typical distance between agents is large) we employ bimolecular reaction kinetic theory \cite{doi1976std,erban2009smr} and basic Markov chain technology \cite{Norris1998} to show how and why the type $\S$ agents are driven to extinction with high probability. Different tools are needed to tackle the crowded high-density regime (when each agent can interact with many others simultaneously); studying the stochastic dynamics of the Fourier modes of the spatial distribution of agents (following \cite{rogers2012dnc,Mckane2014}), and exploiting a separation of time-scales between spatial fluctuations and the ratio of agent types, we are able to demonstrate a systematic bias in favour of type $\W$ agents. In both cases our results match those of stochastic simulations. 

The remainder of the manuscript is organised as follows: after giving a detailed specification of the model and highlighting the essential paradox in the unequal outcomes of simulations, we explore the low- and high-density regimes in detail. Finally, we discuss the wider implications of our work, interpreted in the context of both opinion dynamics and evolutionary biology. 

\section{Model specification}\label{secton:model_specification}
We consider a population of $N$ agents (of types $\S$ and $\W$) that diffuse around a regular $d-$dimensional hypertoroidal domain of length $L$ according to independent Brownian motions with diffusion constant $D$. Each agent attempts to convert agents of the opposite type when they are situated within the focal agent's conversion radius. We write $r_\s,r_\w$ for the conversion radii of the two types and $\lambda_\s,\lambda_\w$ for the rates of conversion (so that, for example, the probability of an $\S$ agent successfully converting a $\W$ agent at distance $x$ in a small period of time $\delta t$ is $\delta t\lambda_\s\mathbb{1}_{|x|<r_\s}$). We are interested in the case that there is a trade-off between the strength of influence (measured by conversion rate) and the region over which it has an effect (measured by conversion radius). In order to mimic limited, but equal resources, we assume the product of the rate of conversion and region of influence is the same for both types, so that 
\begin{equation}
\lambda_\s r_\s^d=\lambda_\w r_\w^d\,.
\label{constraint}
\end{equation}
We choose the type labels $\S$ and $\W$ so that $r_\s\leq r_\w$ and hence $\lambda_\s\geq\lambda_\w$, that is, $\S$-type agents have small conversion radii, but convert strongly and $\W$-type agents have wide conversion radii, but convert weakly. 

Consider a situation in which the agents are distributed uniformly at random in space. In this case each type $\W$ agent has an average of $N_\s L^{-d}v_dr_\w^d$ type $\S$ agents within its conversion radius, where $N_\s$ is total number of $\S$ agents, $L$ is the length of the domain and $v_d=\pi^{d/2}/\Gamma(1+d/2)$ is the volume of the unit $d$-sphere. The expected total rate of conversion from $\S$ to $\W$ is therefore $N_\w N_\s L^{-d}v_d\lambda_\w r_\w^d$. Performing the analogous computation for type $\S$ agents converting type $\W$, we find the expected total rate to be $N_\s N_\w L^{-d}v_d\lambda_\s r_\s^d$. Thus we see that the constraint \eqref{constraint} forces equality between the types in the sense that expected rates of conversion in either direction are equal (at least when the agents are uniformly mixed in space). 

\begin{figure}
\begin{center}
\includegraphics[width=250pt, trim=70 90 50 50, clip=true]{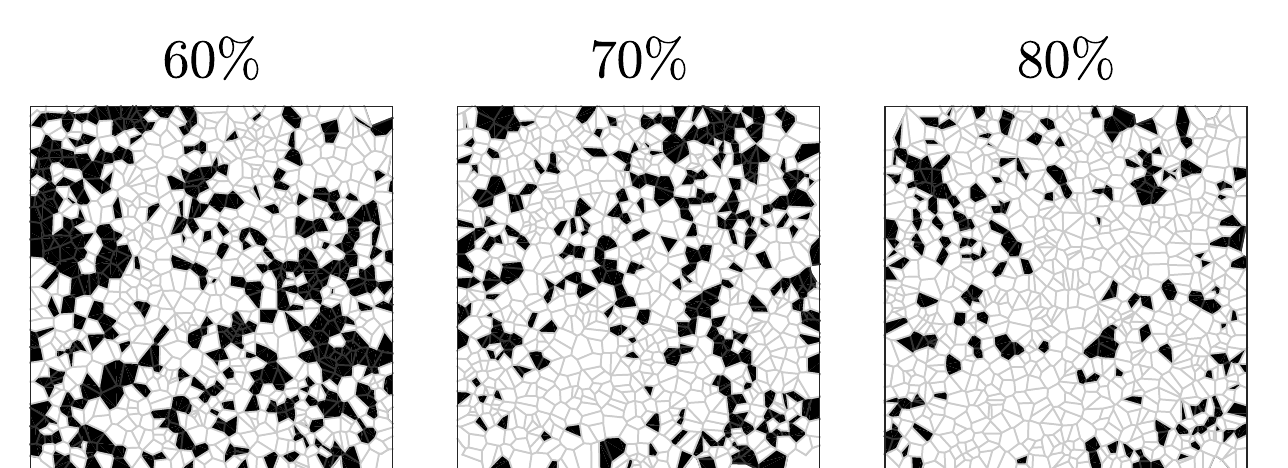}
\end{center}
\caption{Snapshots of the evolution of a single realisation taken at different fractions of $W$-type agents. $N=1000$ individuals move with diffusion coefficient $D=0.01$ on a domain of size $L=1$. For visualisation purposes we have partitioned space into Voronoi cells with the particle positions as seeds. The cells are coloured white for $W$-type agents and black for $S$-type agents. Other parameters are $r_\s=0.1$,$\lambda_\s=1$, $r_\w=0.2$, $\lambda_\w=0.25$, d=2. Note that, unlike the usual 2D voter model, our model does not appear exhibit spatially organised coarsening dynamics.}
\label{snap}
\end{figure}

In simulations, however, we repeatedly observe the wide and weak type $\W$ agents taking over the population. Paradoxically, snapshots of the population during this process do not \emph{to the eye} exhibit any particular spatial ordering (see Fig.~\ref{snap}), making it difficult to see why the above calculation for uniformly mixed agents should not hold. In the next two sections we will look in detail at the dynamics of the model in different regimes to explain this phenomenon. For simplicity we perform our calculations in the case $d=1$, however, the techniques and conclusions are valid for any dimension. 

\section{Low-density regime}\label{section:low_density_regime}
First we consider a 1-D domain of length $L$ that is much larger than the total reactive region $2N_\w r_\w+2N_\s r_\s$. As the agents diffuse in this low-density regime, we expect that, almost always, they interact only in pairs, since the probability of three agents' locations converging is smaller by a factor of $O(r/L)$, where $r$ is the typical size of a conversion radius. We will show that, during these pairwise interactions, the wide and weak type $\W$ agents are in fact more likely to succeed in converting the small and strong type $\S$ agents than vice versa. One intuition for this result comes from the observation that although the exposure of each agent to each other is equal over long times, the type $\W$ agents have the opportunity to convert the type $\S$ agent before it comes within range, giving them a ``head start" in some sense.

We write $\kappa\ll1$ for the rate of occurrence of pairwise encounters that end with one agent converting the other. This rate is a complicated function of the current composition of the population and the locations of the agents. We will see, however, that it is possible to predict the the ultimate fate of the system without explicitly knowing these details. Our strategy is to consider a single `tagged' agent, and examine the density $u(x)$ of agents of the opposite type a distance $x$ from the focal agent, for at small distances $x\in[0,r_\w)$. Making a quasi-stationary approximation that fixes $\kappa$ as a small constant, we solve the equilibrium problem
\begin{equation}
0=2Du''(x)-\lambda(x)u(x)\,,
\label{ueq}
\end{equation}
where $\lambda(x)=\lambda_\w \mathbb{1}_{\{x<r_\w\}}+\lambda_\s \mathbb{1}_{\{x<r_\s\}}$, and with boundary conditions
\begin{equation}
u'(0)=0\,,\quad u'(r_\w)=\kappa\,.
\end{equation}
The fraction of reacting pairs in which the agent with the smaller conversion radius wins is then found to be  
\begin{equation}
p_\s=\frac{u'(r_\s)}{u'(r_\w)}\frac{\lambda_\s}{\lambda_\s+\lambda_\w}\,.
\label{ps}
\end{equation}
To understand this equation, recall that the $u'(x)$ describes the probability flux at $x$, and hence the ratio of fluxes $u'(r_\s)/u'(r_\w)$ gives the fraction of reactions occurring inside the smaller conversion radius. The second term on the right of (\ref{ps}) is simply the probability of the $S+W\rightarrow2S$ reaction occurring before the alternative $S+W\rightarrow2W$, given that the agents have distance less than $r_\s$.
\begin{figure}[t]
\hspace{5mm}\includegraphics[height=140pt, trim=0 0 30 0]{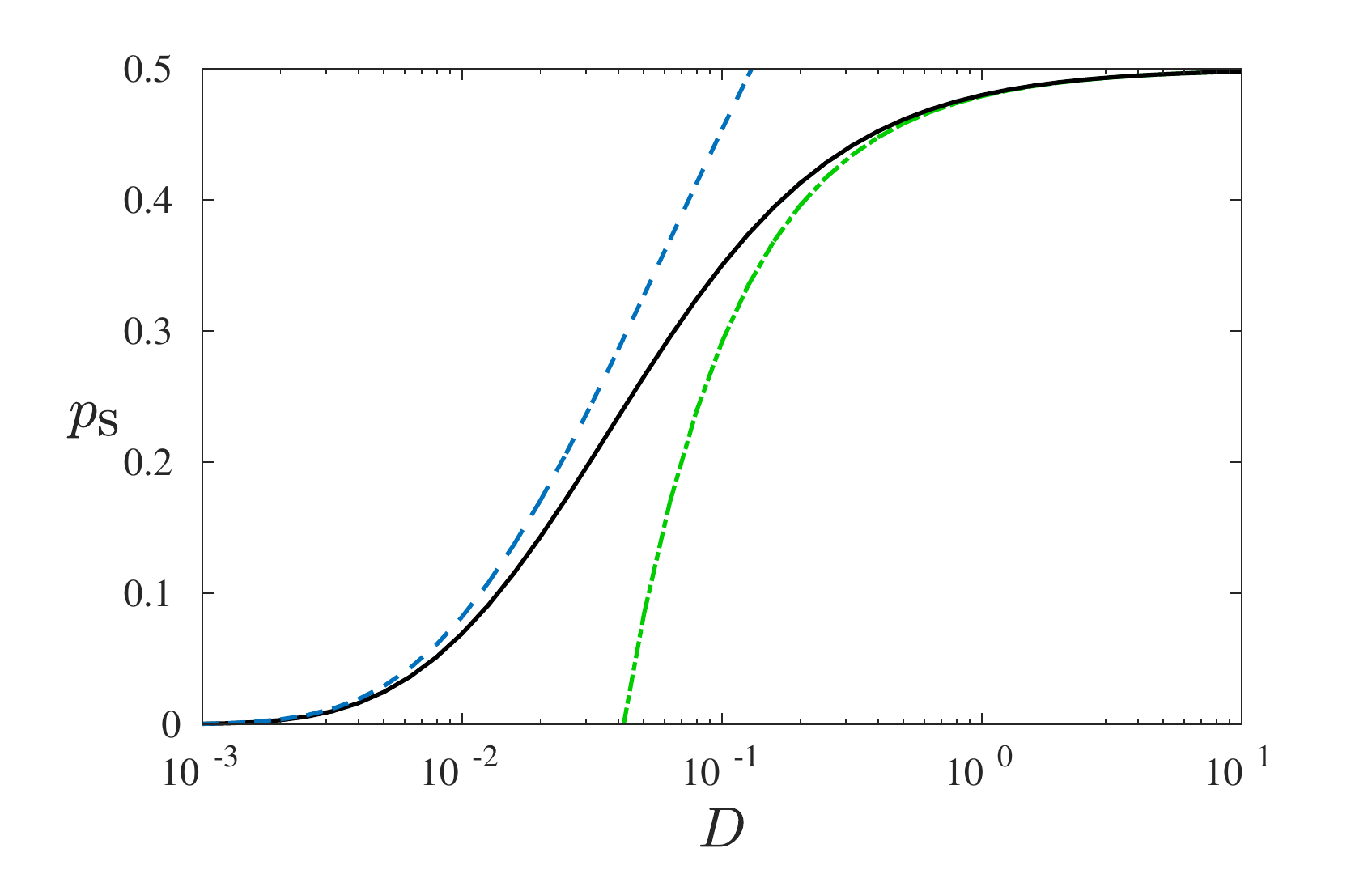}
\caption{Probability of type $\S$ converting type $\W$, as a function of diffusion rate $D$, for $r_\w=1,\,r_\s=1/2$. The solid line is the result of equation \eqref{psfull}, the dashed and dot-dashed lines are the asymptotic results given in \eqref{asymp}.}
\label{Dps}
\end{figure}
\begin{widetext}
\begin{equation}
p_\s=\frac{2 \mu_1 r_\w \left(e^{2 \mu_1r_\s}-1\right) e^{\mu_2(r_\s+r_\w)}}{(r_\s+r_\w) \left((\mu_1-\mu_2)(e^{2 r_\s (\mu_1+\mu_2)}- e^{2 \mu_2 r_\w})+(\mu_1+\mu_2) (e^{2 \mu_1r_\s+2 \mu_2r_\w}-e^{2 \mu_2r_\s})\right)}\,.
\label{psfull}
\end{equation}
\end{widetext}

For our particular problem, since $\lambda(x)$ is piece-wise constant, the solution to \eqref{ueq} may be found by matching solutions to the autonomous problem $0=2Du''(x)-\lambda u(x)$ in the two regions $[0,r_s)$ and $[r_s,r_w)$ with different values of $\lambda$. Introducing $\mu_1=\sqrt{(\lambda_\s+\lambda_\w)/2D}$ and $\mu_2=\sqrt{\lambda_\w/2D}$, and using the boundary condition $u'(0)=0$, we obtain
\begin{equation*}
u(x)=\begin{cases}C_1 \cosh\left(\mu_1 x\right) &\textrm{for } x\in[0,r_\s)\\
C_2 \cosh\left(\mu_2 x\right)+C_3 \sinh\left(\mu_2 x\right) &\textrm{for } x\in[r_s,r_w).\end{cases}
\end{equation*}
The constants $C_{1,2,3}$ above are determined by the requirements of matching density $u(x)$ and flux $u'(x)$ at the interface between regions at $x=r_\s$, as well as the boundary condition $u'(r_\w)=\kappa$. The full expressions are long and largely uninformative. 
Applying these results to \eqref{ps} yields the probability of the agent with the smaller interaction radius agent winning in a pairwise interaction, given in equation \eqref{psfull} overleaf.

Note that $p_\s<1/2$ for $r_\s<r_\w$ and for all $D$, meaning that, in pairwise interactions, the agents with a wide region of influence always have an advantage. The strength of this advantage varies with the difference in radii and the rate of diffusion. The effect of varying the diffusion constant is most clearly demonstrated by considering the asymptotic behaviour of equation \eqref{psfull} for small and $D$. Since $\mu_{1,2}$ diverge for small $D$, we find that $p_\s\to0$ exponentially fast in $D^{-1/2}$; the dominant scaling in this limit being controlled by the leading exponent in \eqref{psfull}. For large $D$, $\mu_{1,2}$ have Taylor series, and a straightforward expansion is possible. Together, we obtain:
\begin{equation}
p_\s\approx\begin{cases}\displaystyle  \frac{2r_w}{r_s+r_w}\exp\left[-(r_\w-r_\s)\sqrt{\frac{\lambda_\w}{2D}}\,\right]\quad&\text{for }D\ll1\,,\\\displaystyle \Bigg.\frac{1}{2}-\frac{r_\w-r_\s}{24D} &\text{for }D\gg1\,. \end{cases}
\label{asymp}
\end{equation}
In both limits it is clear that the advantage held by the $\W$ agents increases with the difference in conversion radii $(r_\w-r_\s)$. These asymptotic scalings are compared to the full expression for $p_\s$ in Figure~\ref{Dps}. 

\begin{figure}[t]
\includegraphics[height=140pt, trim=0 0 30 0]{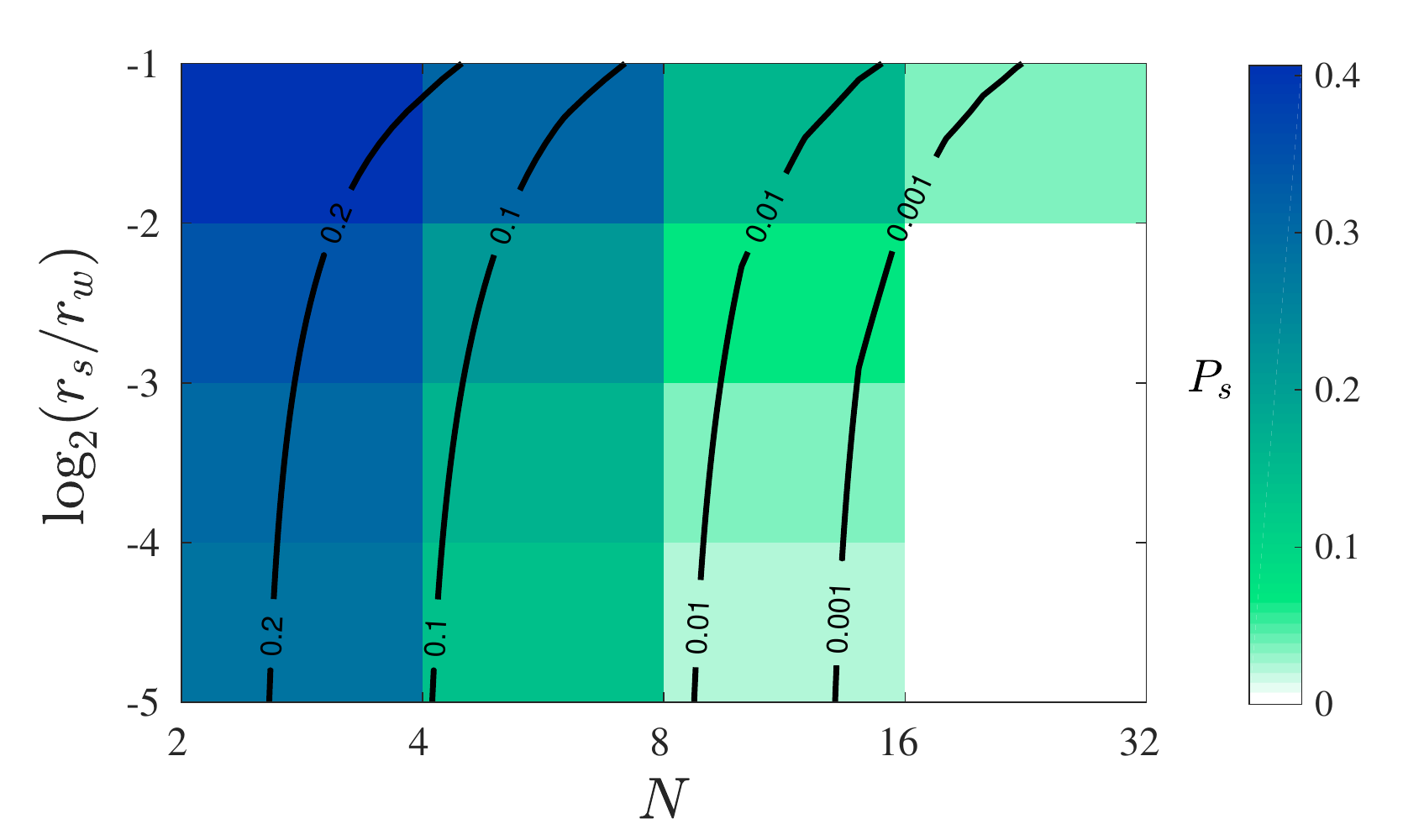}
\caption{Colour plot of the probability $P_\s$ of type $\S$ agents achieving consensus obtained from simulations with various values of population size $N$ and size ratios $r_\s/r_\w$. The overlaid black lines show level contours obtained from \eqref{Ps}; we note that the value of $P_\s$ observed in stochastic simulations is generally slightly higher than the analytical prediction, but the qualitative features match well.}
\label{contour}
\end{figure}

So far we have only discussed the fraction of pairwise interactions won by either agent; for a strategy to dominate the whole population, its agents must win repeatedly. Continuing with the approximation that only pairwise encounters are possible (which is exact in the limit of low density), the number $N_\s$ of S agents (or equivalently $N_\w=N-N_\s$ for W agents) in the population at time $t_i$ of the $i$-th conversion is described by the simple birth-death process
\begin{equation}
\mathbb{P}\big[N_\s(t_i+1)=n\big|N_\s(t_i)=m\big]=\begin{cases}p_\s &\text{if }n=m+1\\1-p_\s&\text{if } n=m-1\,.\end{cases}
\end{equation}
Following standard methods for one-step Markov chains (see, e.g. \cite{Norris1998}), we find that the probability of S agents winning overall, starting from an initially equal mix $N_\s=N_\w=N/2$ is given by
\begin{equation}
P_\s=\frac{p_\s^{N/2}}{p_\s^{N/2}+(1-p_\s)^{N/2}}\,.
\label{Ps}
\end{equation}
Since, for $r_s<r_w$, $p_\s<1/2$ it is straightforward to see that $P_\s\to0$ as $N\to\infty$, meaning that the advantage of the wide agents in pairwise interactions, no matter how small, is amplified into certain victory in large populations. Figure~\ref{contour} shows a comparison between the theoretical result of \eqref{Ps} and the observed win ratio in stochastic simulations for various parameter values.

\section{High-density regime}\label{section:high_density_regime}
The analysis in the previous section hinged on the fact that, in low density populations, agents typically meet only in pairs. For some applications it may be instructive to consider the opposite limit in which we send the population $N\to\infty$, keeping the domain size fixed. This regime requires a completely different treatment, focusing on the spatial distribution of agents.

We first introduce some notation to keep track of the agent locations and types. In a population of size $N$ we index the agents by integers $i\in\mathcal{N}=\{1,\ldots,N\}$, and write $X_i$ for the location agent $i$. We also write $\mathcal{S}$ for the set of indices of S agents, and similarly $\mathcal{W}$ for those of type W, so that $\mathcal{N}=\mathcal{S}\cup\mathcal{W}$ and $\mathcal{S}\cap\mathcal{W}=\varnothing$. The state of the system at some instant in time is then completely specified by the population densities of the type S and W agents, respectively defined as
\begin{equation}
\varphi^\s(x)=\frac{1}{N}\sum_{i\in \mathcal{S}}\delta(x-X_i)\,,\quad \varphi^\w(x)=\frac{1}{N}\sum_{i\in \mathcal{W}}\delta(x-X_i)\,.
\end{equation}

According to the model definition we gave in the introduction, the state of the system evolves via the continuous diffusion of the agents, and the discrete jumps occurring when one agent converts another. For technical reasons, it is simpler in this section for us to express the diffusion of agents as the limit of a position-jump process. In this description all agents are instantaneously static, but may choose to jump at random times that are exponentially distributed with rate $\gamma$; when a jump occurs the distance travelled is a normal random variable with mean zero and variance $D/\gamma$. In the limit $\gamma\to\infty$ the agents' trajectories converge to Brownian motion with diffusion coefficient $D$. 

This description is useful as now the state of the system (i.e. the pair $\bm{\varphi}=(\varphi^\s,\varphi^\w)$) evolves according to a Markov jump process in the space of distributions. Changes to the system state (whether due to an agent moving, or changing type) are expressed as the subtraction of a Dirac mass from one or other density, combined with addition of another Dirac mass, possibly in a spatially distinct location. Following the notation of \cite{Rogers2012} we introduce operators $\Delta^{\u\pm}_x$ to denote the addition ($+$) or subtraction ($-$) of a Dirac mass of weight $1/N$ at spatial location $x$ in the density of agents of type $\u$. Writing $P(\bm{\varphi},t)$ for the probability density of state $(\bm{\varphi})$ at time $t$, we are able to formulate the master equation
\begin{equation}
\frac{\d}{\d t}P(\bm{\varphi};t)=N\iint \mathcal{Q}(\bm{\varphi};x,y) P(\bm{\varphi};t)\,\d x\,\d y\,,
\label{master}
\end{equation}
where 
\begin{equation}
\begin{split}
\mathcal{Q}(\bm{\varphi};x,y)=&\big(\Delta_x^{\s+}\Delta_y^{\s-}-1\big)\,\varphi^\s(x)\,d(x-y)\\
+&\big(\Delta_x^{\w+}\Delta_y^{\w-}-1\big)\,\varphi^\w(x)\,d(x-y)\\
+&\big(\Delta_x^{\s+}\Delta_x^{\w-}-1\big)\,\varphi^\s(x)\,\varphi^\w(y)\,\lambda^\w(x-y)\\
+&\big(\Delta_x^{\w+}\Delta_x^{\s-}-1\big)\,\varphi^\w(x)\,\varphi^\s(y)\,\lambda^\s(x-y)\,.
\end{split}
\label{defQ}
\end{equation}
The four terms above respectively correspond to: movement of an S agent, movement of a W agent, conversion of an S agent to  a W, conversion of a W agent to an S. The functions $d$, $\lambda^\s$ and $\lambda^\w$ are the rate kernels for these events in agents of different types; they are given by
\begin{equation}
\begin{split}
&d(x)=\frac{\gamma}{\sqrt{2\pi D/\gamma}}e^{-\gamma x^2/2D}\,,\\
&\lambda^\s(x)=\lambda_\s\mathbb{1}_{|x|<r_\s}\,,\quad \lambda^\w(x)=\lambda_\w\mathbb{1}_{|x|<r_\w}\,,\bigg.
\end{split}
\end{equation}
although the theory holds for alternative forms of movement and interaction.
Equations \eqref{master} and \eqref{defQ} define a Markov jump process in the space of degenerate distributions composed of multiple Dirac masses. In the high-density limit of large $N$, as more and more agents are included in the system, we expect that the normalised agent densities $\bm{\varphi}$ will converge (in some suitable sense) to smooth random functions. To bridge the gap between the discrete finite populations and the smooth large $N$ approximation, it is useful to regularise the densities by moving to Fourier space. We introduce the series expansions 
\begin{equation}
f(x)=\sum_{k\in\mathbb{Z}}f_k\,e^{ikx}\,,
\end{equation}
for $f\in\{\varphi^\s,\varphi^\w,\lambda^\s,\lambda^\w,d\}$, where $f_k=\frac{1}{2\pi}\int e^{-ikx}f(x)\,\d x\,.$ In Fourier space, the step operators for addition or subtraction of Dirac masses can themselves be expressed as a series in large $N$; 
\begin{equation}
\begin{split}
\Delta_x^{\u\pm}=1&\pm\frac{1}{2N\pi}\sum_ke^{-ikx}\frac{\partial}{\partial \varphi^{\u}_k}\\&+\frac{1}{8N^2\pi^2}\sum_{k,\ell}e^{-i(k+\ell)x}\frac{\partial}{\partial \varphi^{\u}_k}\frac{\partial}{\partial \varphi^{\u}_\ell}+\mathcal{O}(N^{-3})\,.
\end{split}
\end{equation}
Inserting these expressions into to \eqref{master} via \eqref{defQ} and discarding terms above second order, we obtain a Fokker-Planck equation for the Fourier modes of $\varphi^\s$ and $\varphi^\w$:
\begin{equation}
\frac{\partial P}{\partial t}=-\sum_{\u,k} \frac{\partial}{\partial \varphi^{\u}_k}A^{\u}_{k}P+\frac{1}{2N}\sum_{\u,\v,k,\ell} \frac{\partial}{\partial \varphi^{\u}_k}\frac{\partial}{\partial \varphi^{\v}_\ell}B^{\u,\v}_{k,\ell}P
\label{fourFPE}
\end{equation}
where, introducing
\begin{equation}
c^\pm_k=\sum_m\Big(\lambda^\w_m\varphi^\w_m\varphi^\s_{k-m}\pm\lambda^\s_m\varphi^\s_m\varphi^\w_{k-m}\Big)\,, 
\end{equation}
and taking the diffusion jump rate $\gamma\to\infty$, we have
\begin{equation}
A^{\w}_{k}=2\pi c^-_k-\frac{D k^2}{2}\varphi^\w_k\,,\quad A^{\s}_{k}=-2\pi c^-_k-\frac{D k^2}{2}\varphi^\s_k\,,
\end{equation}
and
\begin{equation}
B^{\u,\u}_{k,\ell}=c^+_{k+\ell} +\frac{D k\ell}{4\pi}\varphi^{\u}_{k+\ell}\,,\quad B^{\u,\v}_{k,\ell}=-c^+_{k+\ell}\,.
\end{equation}
From the Fokker-Planck equation \eqref{fourFPE} we can extract statistical information about the system behaviour. Taking the limit $N\to\infty$ the dynamics become non-random; multiplying though by either $s_k$ or $w_k$ and integrating over all modes yields, in this limit, the mean-field ODEs
\begin{equation}
\frac{\d \varphi^\w_k}{\d t} = 2\pi c^-_k-\frac{D k^2}{2} \varphi^\w_k\,,\quad \frac{\d \varphi^\s_k}{\d t} = -2\pi c^-_k-\frac{D k^2}{2} \varphi^\s_k  \,.
\label{mf}
\end{equation}
It is easy to check that this system admits a family of fixed points of the form
\begin{equation}
\varphi^\w_k=\frac{\omega}{2\pi}\delta_{k,0}\,,\quad\varphi^\s_k=\frac{1-\omega}{2\pi}\delta_{k,0}\,,\quad \omega\in[0,1]\,.
\label{fix}
\end{equation}
These correspond to the situation in which the agents are homogeneously mixed in space, with a fraction $\omega$ being of type W and $(1-\omega)$ of type S. These fixed points are linearly stable with respect to perturbations in modes $k\neq 0$, but perturbations to $\varphi^\w_0,\varphi^\s_0$ are not suppressed. For large but finite $N$, the noisy behaviour of the system gives rise to small fluctuations in all modes, which may in turn induce a net drift in the number of agents of each type. 

To explore this possibility, we propose the linear fluctuation ansatz\footnote{Note this is not quite the same as the Van Kampen expansion, since we preserve the order $1/N$ contribution to the dynamics of $\omega$.}
\begin{equation}
\varphi^\w_k=\frac{\omega}{2\pi}\delta_{k,0}+\frac{\xi_k}{\sqrt{N}}\,,\quad\varphi^\s_k=\frac{1-\omega}{2\pi}\delta_{k,0}-\frac{\xi_k}{\sqrt{N}}\,,
\end{equation}
which gives
\begin{equation}
\begin{split}
c^-_k=&\frac{\omega(\lambda^\s_k-\lambda^\w_0)+(1-\omega)(\lambda^\w_k-\lambda^\s_0)}{2\pi\sqrt{N}}\xi_k\\&\hspace{20mm}+\frac{1}{N}\sum_m(\lambda^\s_m-\lambda^\w_m)\xi_{m}\xi_{k-m}\,. \\
c^+_k=&\frac{\omega(1-\omega)(\lambda^\w_0+\lambda^\s_0)}{4\pi^2}\delta_{k,0}+\mathcal{O}(1/\sqrt{N})\,.
\end{split}
\label{fsatz}
\end{equation}
Multiplying through by $\omega=2\pi\varphi^\s_0$ and integrating in the Fokker-Planck equation \eqref{fourFPE} (and recalling that, since $\varphi^\w,\varphi^\s\in\mathbb{R}$, we must have $\xi_{-k}=\overline{\xi_k}$), we find 
\begin{equation}
\frac{\d\omega}{\d t}=\frac{4\pi^2}{N}\sum_k(\lambda^\s_k-\lambda^\w_k)\langle|\xi_{k}|^2\rangle\,,
\label{domega}
\end{equation}
where $\langle\cdots\rangle=\int_\mathbb{C} (\cdots)P(\bm{\varphi},t)\prod_{k,\ell\in\mathbb{Z}}\d\varphi^\s_k\d\varphi^\w_\ell$ denotes the average over $P$. The pre-factor of $N^{-1}$ here shows that the two agent types are equally balanced in the limit $N\to\infty$, where $\dot \omega\to0$ and the fraction of agents of each type does not change. Nonetheless, we can also see clearly from this equation that the variance of the fluctuations $\xi_k$ gives rise to a bias of order $N^{-1}$ in the dynamics of $\omega$, with the cumulative effect of slowly but reliably altering the fraction of agents of each type. 

\begin{figure}
\begin{center}
\includegraphics[height=140pt]{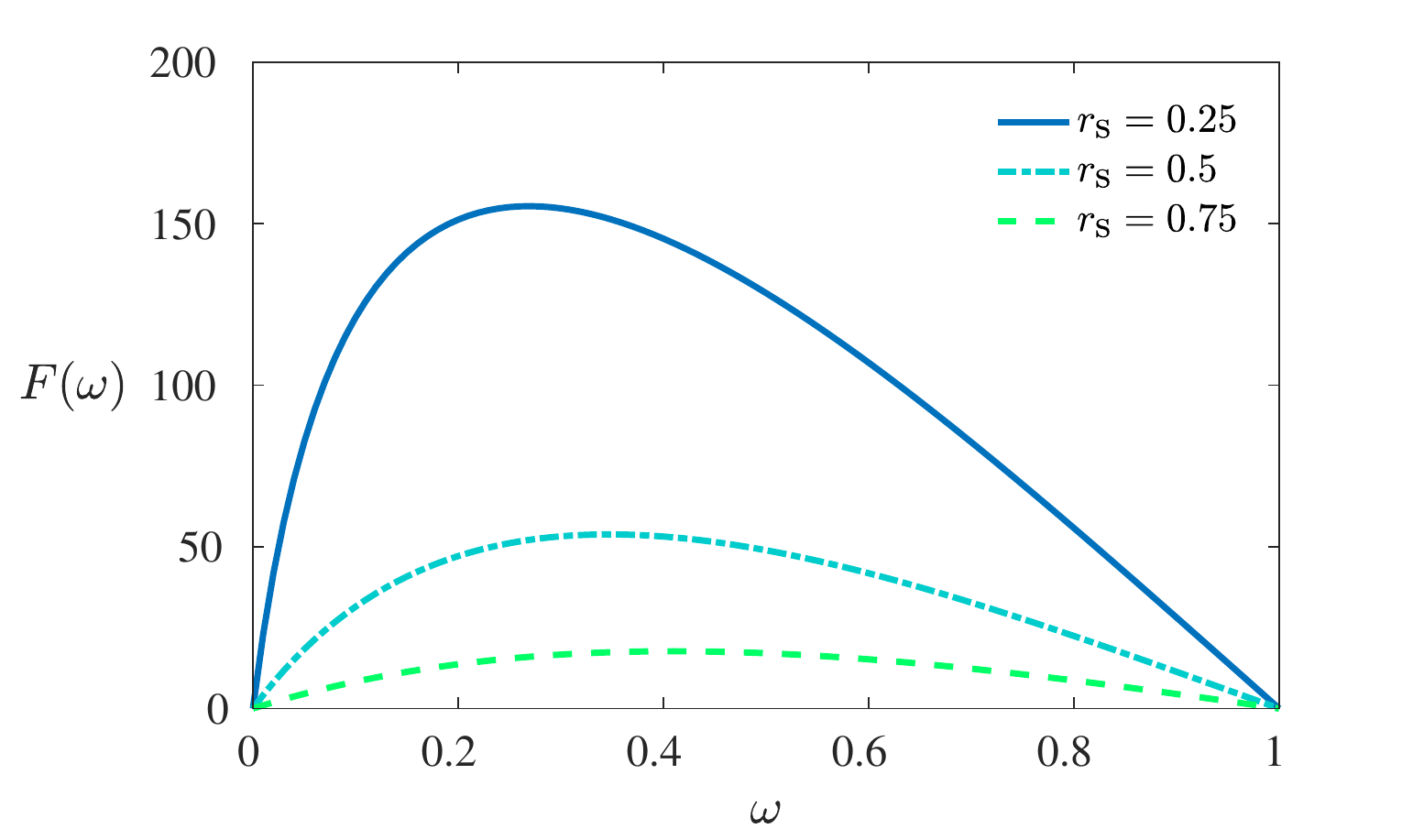}
\end{center}
\caption{Illustration of $F(\omega)$ defined in \eqref{Feq}, for different values of $r_\s$, when $r_\w=1$, $D=0$.}
\label{dtaudt}
\end{figure}

\begin{figure}
\begin{center}
\includegraphics[height=140pt]{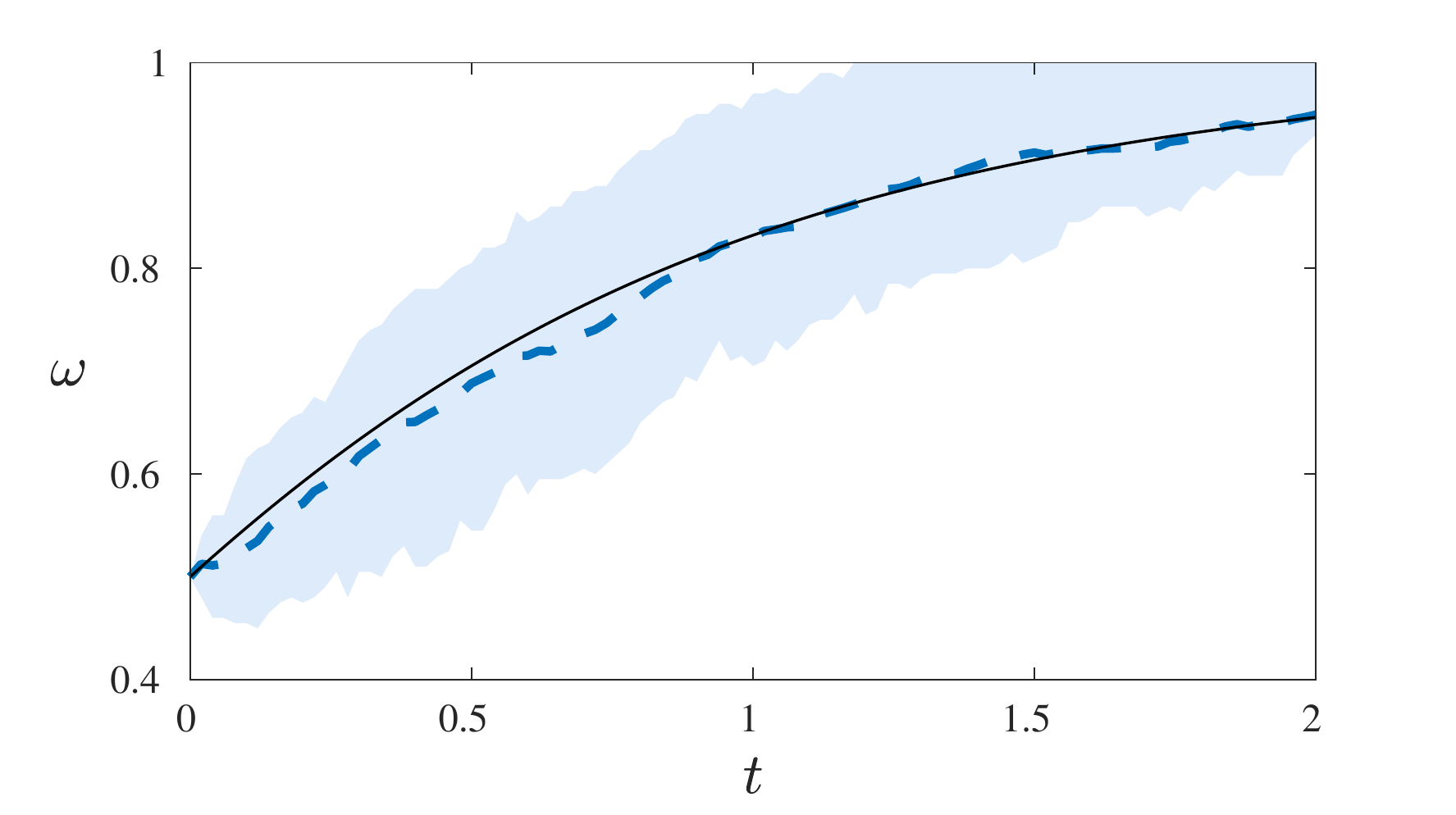}
\end{center}
\caption{The fraction $\omega$ of type W agents increases over time in the high-density regime, in the domain $(-\pi,\pi]$ with the diffusion coefficient set to zero. The black curve shows the theoretical prediction \eqref{domegadt} for the case $r_\w=1, r_\s=1/2$. The thick dashed blue line shows the average over 100 simulations in a population of size $N=100$ with the same parameters. The area between the 0.2 and 0.8 quantiles is shaded light blue.}
\label{t_omega}
\end{figure}

To compute the direction and strength of this effect, we multiply through by $\xi_k\xi_{-k}$ in \eqref{fourFPE}, using the ansatz \eqref{fsatz}, and integrate to obtain
\begin{equation}
\frac{\d}{\d t}\langle|\xi_k|^2\rangle=-\alpha_k(\omega)\langle|\xi_k|^2\rangle+\beta_k(\omega)+\mathcal{O}(N^{-1/2})\,,
\label{dxi}
\end{equation}
where
\begin{equation}
\begin{split}
\alpha_k(\omega)&=\frac{\omega(\lambda^\s_0-\lambda^\w_k)+(1-\omega)(\lambda^\w_0-\lambda^\s_k)}{\pi}+Dk^2\,,\\
\beta_k(\omega)&=\frac{\omega(1-\omega)}{2\pi^2}\,.
\end{split}
\end{equation}
If we neglect the $\mathcal{O}(N^{-1/2})$ terms in \eqref{dxi} then that equation, together with \eqref{domega}, specifies a closed system of ODEs. Moreover, the pre-factor of $N^{-1}$ in the right hand side of \eqref{domega} implies a separation of time-scales between the fast dynamics of the fluctuations and slow dynamics of the overall ratio of types. The system can be solved, to leading order in $N$, via adiabatic elimination (see e.g. \cite{Gardiner1985,Parsons2017} for discussion of this and related methods). Treating $\omega$ as a constant in \eqref{dxi}, then for $1\ll t\ll N$ we have 
\begin{equation}
\langle|\xi_k|^2\rangle \approx \frac{1}{2\pi} \frac{\omega(1-\omega)}{\omega(1-\lambda^\w_k)+(1-\omega)(1-\lambda^\s_k)+\pi Dk^2}\,.
\label{xilim}
\end{equation}
The time-scale of relaxation to this quasi-equilibrium is much faster than the dynamics of $\omega$, so we may treat the variance of the fluctuations $\langle |\xi_k|^2\rangle$ as fully determined by the value of $\omega$. Inserting \eqref{xilim} into \eqref{domega}, we obtain the differential equation
\begin{equation}
\begin{split}
\frac{\d\omega}{\d t}&=\frac{1}{N}F(\omega)\,,
\end{split}
\label{domegadt}
\end{equation}
where 
\begin{equation}
F(\omega)=\sum_k\frac{2\pi(\lambda^\s_k-\lambda^\w_k)\omega(1-\omega)}{\omega(\lambda^\s_0-\lambda^\w_k)+(1-\omega)(\lambda^\w_0-\lambda^\s_k)+\pi Dk^2}\,.
\label{Feq}
\end{equation}
Figure~\ref{dtaudt} illustrates the shape of this function for several different values of $r_\s$. Note that in each case $F(\omega)$ is strictly positive over the whole of $\omega\in(0,1)$, meaning that the trend is always for the fraction of type $\W$ agents to increase on average. In Figure~\ref{t_omega}, we show an example evolution trajectory of $\omega$, given by equation \eqref{domegadt}, compared to the average over a number of simulation runs, which shows excellent agreement. 

\section{Conclusion} We have introduced a novel adaptation of the voter model in which the two types of agents are mobile and differentially persuasive. As we have demonstrated, our model displays the interesting phenomenon that, despite being equally persuasive on average, the agents that distribute their resources widely at the cost of converting others only weakly have an advantage over the forceful but narrowly-focussed individuals. Our purpose here has been to investigate this surprising behaviour in an analytically tractable setting, however, we are hopeful that our results might form the basis of new investigations into spatial consensus-forming dynamics, inspired by applications to political campaigning. 

To explore if the effect we have observed in the highly-stylised model presented here is relevant to real-world applications, it will be necessary to consider generalisations of several aspects of the model. The results presented in this letter are for the simplest possible conversion kernels and for agents diffusing at the same rate and in one spatial dimension. First we note that extension to higher spatial dimensions is relatively straightforward: in the low-density regime one should modify equation (\ref{ueq}) to employ the appropriate Laplacian operator for diffusion, then the analysis produces as in the 1D case; in the high-density regime it is necessary to take a Fourier transform in each spatial co-ordinate, but otherwise the technique and results are unchanged. In real scenarios it is possible that there is a trade-off between campaigning strategy (encoded by the conversion kernel) and speed of movement. It is also likely that the resources of the two types may not be equal and it is of interest to characterise the scenarios in which the wide and weak campaigners dominate despite being more poorly resourced.  It may also be that the space through which the agent's diffuse is not physical space, but rather the space of connections formed by a social network, for example. Since  the spread of memes on real social networks is well studied \cite{chen2011ims,lerman2010ice}. It may be possible, therefore, to validate or falsify the model against real data.

We also anticipate that the results of this letter may inspire interesting investigations beyond the realms of political campaigning, for example to biological competition \cite{Reichenbach2007}. Indeed, there are clear parallels between this work and that of \cite{pigolotti2014sad}, who demonstrated that faster diffusing individuals dominate in spatial models of biological competition, and \cite{Galla2017} who examine the evolutionary dynamics of mobile agents in chaotic flows. The underlying competition dynamics and mechanisms of establishing dominance are, however, entirely different. Another context for future research, building on our results, is in the interaction between the dynamics of individual behaviour and disease spread; in rabies, for example, infected individuals exhibit markedly different behaviours to the uninfected \cite{Murray1986}. There are also consequences for the understanding of the evolution of modes of infection, for example the evolutionary trade-offs between long-range but unreliable air-borne transmission to blood-borne infections requiring personal contact. 

\section{Acknowledgements} 
KM is supported by the Bank of Butterfield. TR is supported by the Royal Society.


\end{document}